\documentclass[titlepage]{article}

\usepackage{graphicx}
\usepackage{booktabs}
\usepackage{url}

\usepackage{abstract}

\begin{document}

\title{The production of information in the attention economy}

\date{}

\author{
  Giovanni Luca Ciampaglia\thanks{To whom correspondence should be addressed.
  Email: gciampag@indiana.edu.}, Alessandro Flammini,
  Filippo Menczer\\
  \\
Center for Complex Networks and Systems Research, \\
School of Informatics and Computing, \\
Indiana University, \\
Bloomington, IN 47408, USA 
}


\renewcommand{\thefootnote}{\fnsymbol{footnote}}
\maketitle
\renewcommand{\thefootnote}{\arabic{footnote}}

  \begin{abstract}

    Online traces of human activity offer novel opportunities to study the
    dynamics of complex knowledge exchange networks, and in particular how the
    relationship between demand and supply of information is mediated by
    competition for our limited individual attention. The emergent patterns of
    collective attention determine what new information is generated and
    consumed. Can we measure the relationship between demand and supply for new
    information about a topic? Here we propose a normalization method to compare
    attention bursts statistics across topics that have an heterogeneous
    distribution of attention. Through analysis of a massive dataset on traffic
    to Wikipedia, we find that the production of new knowledge is associated to
    significant shifts of collective attention, which we take as a proxy for its
    demand. What we observe is consistent with a scenario in which the
    allocation of attention toward a topic stimulates the demand for information
    about it, and in turn the supply of further novel information. Our attempt
    to quantify demand and supply of information, and our finding about their
    temporal ordering, may lead to the development of the fundamental laws of
    the attention economy, and a better understanding of the social exchange of
    knowledge in online and offline information networks.

  \end{abstract}




Massive logs on online human activity create new possibilities to study complex
socio-economic phenomena\cite{Crane2008, Moat2013, Miritello2013}. Among these,
the dynamics of knowledge exchange networks, and in particular the emergent
interactions between producers and consumers of information\cite{Weng2013},
have not been explored like the flows of material goods. Yet they have a
critical impact on our opinions, decisions, and lives\cite{Benkler2006}. 

An overwhelming amount of information stimuli compete for our
cognitive resources, giving rise to the \emph{economy of attention}
\cite{Shapiro1999}, first theorized by Simon \cite{Simon1971}. At the aggregate
level, this phenomenon is often referred to as \emph{collective attention}. Work
on collective attention has mainly focused on the \emph{consumption} of
information \cite{Wu2007, Hodas2012, Weng2012}. Characteristic signatures of
information consumption have been shown to correlate with real-world events,
such as the spread of influenza \cite{Ginsberg2009}, financial stock returns
\cite{Moat2013}, and box office results \cite{Mestyan2013}. 

The \emph{production} side of the equation --- whether and how the creation of
information is driven by demand --- has been explored to a limited extent in the
literature, owing in part to the challenges in quantifying information demand.
Imitation of popular content \cite{Leskovec2009}, for instance, is the simplest
form of supply matching demand for information. However, while examples of
imitation of online contents abound \cite{Simmons2011}, they do not point to a
quantitative relationship between the demand for and production of information.
In looking at the role of attention as a possible driver for the generation of
novel content, Huberman \textit{et al}. found a positive correlation between the
productivity of YouTube contributors and the number of views of their previous
videos \cite{Huberman2009}. This confirms that prestige is a powerful motivation
for creation of knowledge \cite{Franck1999}. 

Here we tackle the measurement of demand and supply of information goods and
their relative ordering in time. Looking at attention toward a specific piece of
information, no link between traffic bursts and the number of edits to a Wikipedia
article has been found so far \cite{Kampf2012}. We focus on the \emph{creation}
of Wikipedia articles as a better proxy for the production of information, and
on visits to topically related articles as a proxy for its demand. Analysis of
Wikipedia traffic data thus allows us to study how the generation of new
knowledge about a topic precedes or follows its demand.

More specifically, we are interested in how attention towards topics changes
around the time that new knowledge about them is created. Moreover, we want to
do so by comparing a broad range of topics. Sudden changes of attention, or
``bursts'', have been traditionally studied using the logarithmic derivative
$\Delta N_t/N_t$, where $N_t$ is the number of visits or links accrued by a
topic (e.g. a Wikipedia page, a YouTube video, etc.) during a fixed sampling
interval $t$ and the numerator is customarily defined as $\Delta N_t = N_t -
N_{t-1}$ \cite{Ratkiewicz2010a,Lehmann2012,Kampf2012}. However, the distribution
of $\Delta N/N$ is known to be broad, with a heavy-tail decay that follows a
power-law distribution\cite{Ratkiewicz2010a}. This lack of a characteristic
scale thus makes it difficult to use $\Delta N/N$ for comparing diverse topics.
Here we propose to use a different measure of traffic change based on a simple
normalization of the traffic, in a way that takes into account this and other
confounding factors, such as traffic seasonality and circadian rhythms of
activity \cite{Goncalves2008,Meiss2008}.

\begin{figure*}[ht] 
  \includegraphics[width=\textwidth]{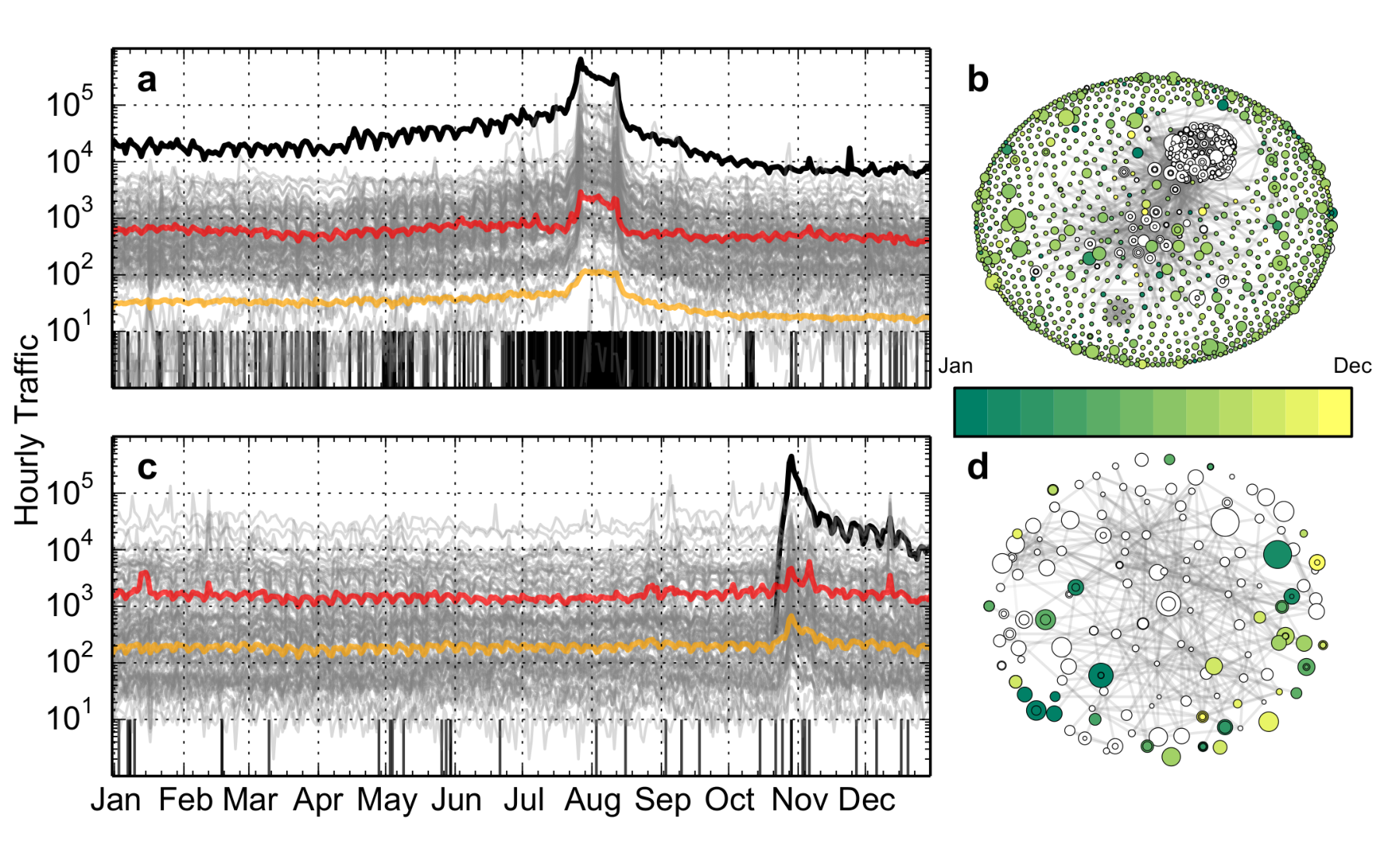}
  \caption{\textbf{Synchronous traffic bursts associate to increased creation
  frequency in two high-profile topics}. \textbf{a},~Time series of traffic.
  The grey lines represent the daily traffic to articles that
  are linked from/to the article ``2012 Summer Olympics,'' according to a recent
  snapshot of Wikipedia (see Methods). For visualization purposes, only a random 
  sample of 100 neighbors is shown. The focal page is represented by the
  black solid line; red and gold lines represent the average and median traffic,
  respectively. The vertical black segments represent the times when new linked
  articles are created (see Methods). \textbf{b},~Network of neighbors of ``2012
  Summer Olympics.'' White nodes represent the neighbor articles predating 2012;
  colored nodes correspond to neighbors created in 2012. The size of the nodes
  is proportional to their yearly traffic volume; their position was computed
  using the ARF layout \cite{Geipel2007}. \textbf{c} and \textbf{d},~Same 
  visualizations as (a) and (b) for the entry about Hurricane Sandy and its neighbors.  
  New articles tend to be peripheral to these networks.}
\label{fig:trafficseries}
\end{figure*}

Wikipedia is currently the fifth most visited Internet website \cite{Alexa.com},
and includes 30 million articles in 287 languages. The English version alone
consists of roughly 4.4 million articles and is consulted, on average, by about
300 million people every day. Each entry, or article, of Wikipedia corresponds
to a separate web page. Wikipedia can thus be regarded as a large information
network, where one can identify broad macroscopic topics. By way of example,
Fig.~\ref{fig:trafficseries} depicts the traffic to two high-profile articles
selected from the 2012 Google Zeitgeist \cite{Google2012} and to their
neighbors. We define a topic as such a page, together with all of its neighbors
--- articles linked by it or linking to it, subsequently to its creation (see
Methods). The networks formed by the two topics are shown in
Fig.~\ref{fig:trafficseries}(b,d).

The volume of traffic to a page or a topic is measured by daily browser requests
for the corresponding pages. Weekly fluctuations are evident in the traffic
patterns shown in Fig.~\ref{fig:trafficseries}(a,c). It is also possible to
observe synchronous bursts of activity, corresponding to increased attention
toward the topic. For the Olympics topic, such increase of attention takes the
form of an anticipatory buildup, leading to two peaks around the opening and
closing ceremonies, followed by a relaxation. For Hurricane Sandy a sudden spike
occurs at the time of creation of the main article, due to the demand of
information about the effects of the hurricane. 

Phenomena like these have been already observed in a wide range of
information-rich environments \cite{Ratkiewicz2010, Ratkiewicz2010a, Crane2008,
Lehmann2012}. During the period of increased attention we observe that new
articles about the Olympic Games are created at a higher frequency. A weaker
pattern is observed for Hurricane Sandy. To quantify the temporal relation
between demand and production of information about a topic, we performed a
systematic study over a large sample of articles. An increase of attention
toward the topic of an article is revealed by an increase in requests for pages
in that topic compared to other topics.

Let us consider a newly created article. A burst of attention for pages related
to it occurring before its creation is consistent with a model in which demand
drives the supply of information. Conversely, a burst that follows its
creation suggests that demand follows supply. On the other hand, if traffic
bursts concomitant with the creation of new articles are no different than
those observed at any other time, then we shall conclude that production and
consumption of information are two unrelated processes.

\section*{Results}

Our analysis is focused on the year 2012. We collected the neighbors of 93,491
pages created during that year. For each created page we considered the two
weeks before and after its creation, and measured the volume of traffic to its
topic in each week. We characterize the typical traffic to the topic in the week
after and before with the median traffic to neighbors $V^{(a)}$ and $V^{(b)}$,
respectively. Let us define infra-week traffic volume change $\Delta V = V^{(a)}
- V^{(b)}$, total volume $V = V^{(a)} + V^{(b)}$, and relative volume change
$\Delta V/V$. 

\begin{figure*}[ht]
  \includegraphics[width=\textwidth]{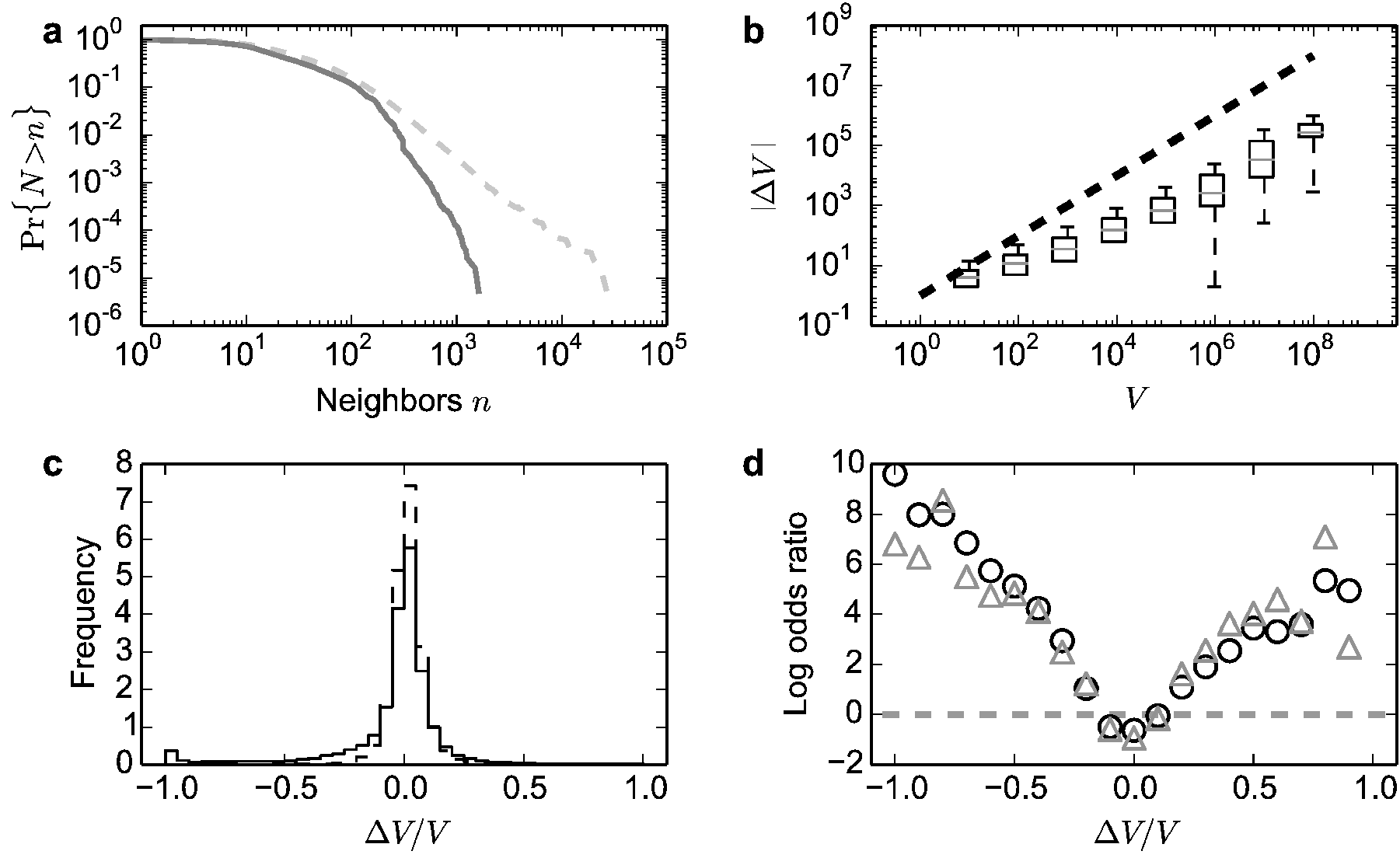} 
  \caption{
    \textbf{Traffic bursts concomitant with creation of new articles differ from
    normal traffic patterns}.
    \textbf{a},~Cumulative distribution of neighborhood size for articles
    created in 2012 (solid), and pre-existing 2012 (dashed). Neighbors are all
    articles linking to or from the focal page. Older articles tend to have
    larger neighborhoods. \textbf{b},~Absolute infra-week traffic change
    $\left|\Delta V\right|$ as a function of total traffic volume $V$ for
    articles created in 2012. Even though some topics may receive hundreds of
    million visits, the change in traffic volume is on average much smaller.
    Pre-existing pages show a very similar pattern. Boxes stretch from the first
    to the third quartile, and whiskers represent the 99\% confidence interval.
    Gray segments within boxes indicate the median. The dashed line is a guide
    to the eye for linear scaling. \textbf{c},~Distribution of the relative
    change in traffic volume for 2012 (solid) and pre-existing (dashed) pages. \textbf{d},~Log
    odds ratio comparing pages created in 2012 versus existing pages as a
    function of relative traffic change, for the whole sample (circles), and for
    a sub-sample of 16,816 pages (18\%) with $V > 2\times 10^5$ visits
  (triangles). The dashed gray line indicates equal odds.}
  \label{fig:change} 
\end{figure*}

For comparison purposes, we collected the neighbors of a roughly equally-sized
sample of pre-existing articles (created before 2012) and analogously computed
their relative infra-week changes in traffic volume over random two-week windows
in 2012. Articles in the baseline sample are older and therefore tend to have
more neighbors, as shown in Fig.~\ref{fig:change}(a). This and other temporal
effects are discounted by considering the relative change in volume $\Delta V/V$
(see Methods). 

We observe that the volume change $|\Delta V|$ scales sublinearly with the total
volume $V$, as illustrated in Fig.~\ref{fig:change}(b). Consequently
$\left|\Delta V\right|/V$ goes to zero as $V$ increases. While the distributions
of $\Delta V/V$ are sharply peaked around zero in both samples
(Fig.~\ref{fig:change}(c)), they are different: a non-parametric
Kolmogorov-Smirnov test rejects the null hypothesis that the two samples of
relative traffic change are drawn from the same distribution ($D = 0.034, p <
0.001$); an Anderson-Darling test, which gives less weight to the median values
of the distribution in favor of the tails, yields similar results. One way to
quantify and interpret the difference between the two distributions is to
compute the ratio of odds that a given change in traffic volume is observed when
a page is created versus when a page has existed for a specific amount of time.
Fig.~\ref{fig:change}(d) plots the log odds as a function of $\Delta V/V$. For
example $\Delta V/V = -0.5$ is over two orders of magnitude more likely to be
observed in a new page compared to a page of generic but fixed age. As shown in
the figure, this effect holds even when we consider only neighborhoods with a
high volume of traffic, which may be indicative of more developed, and hence
more popular topics. In summary, while we find both instances in which bursts in
demand precede ($\Delta V/V < 0$) and follow ($\Delta V/V > 0$) the generation
of new knowledge, comparison with the baseline yields a significant shift
towards the former case, suggesting that consumption anticipates the production
of information.

Which kinds of articles precede or follow demand for information? In
Table~\ref{tab:topbursts} we list a few articles with the largest positive and
negative bursts. Topics that precede demand ($\Delta V/V > 0$) tend to be about
current and possibly unexpected events, such as a military operation in the
Middle East and the killing of the U.S. Ambassador in Libya. These articles are
created almost instantaneously with the event, to meet the subsequent demand.
Articles that follow demand ($\Delta V/V < 0$) tend to be created in the context
of topics that already attract significant attention, such as elections, sport
competitions, and anniversaries. For example, the page about Titanic survivor
Rhoda Abbott was created in the wake of the 100th anniversary of the sinking.

\section*{Discussion}

Our result shows that in many cases, demand for information precedes its supply.
We propose a model to interpret this finding, analogous to the law of supply and
demand \cite{Mas-Colell1995}. An increase in demand indicates a willingness to
pay a higher price for a physical good, which in turn leads to an increase in
supply. In the domain of information, attention plays the role of price: an
increase in demand for information about a topic indicates a higher attention
toward that topic, which in turns leads to the generation of additional
information about it. This model predicts a causal link between demand and
supply of information. Our empirical observations are consistent with this 
prediction, and may represent a first step toward the development of
the fundamental laws of the attention economy.

Of course, not all requests are generated as a result of demand for information.
A number of requests to related articles are likely to be generated by the very
creators of new entries; one could hardly create new knowledge about some
topic without consulting existing pages about it. This is a source of potential
bias for our measure of demand especially in the case of low-traffic topics,
such as entries about small towns or niche musical bands. On the other hand,
significant bursts in volume are observed for popular topics as well (cf.
Figure~\ref{fig:change}(d)). Such bursts could not possibly be generated by the
activity of contributors, who are a small percentage of the Wikipedia audience
\cite{WikimediaFoundation2014}.

As a practical consequence of our finding, volumetric data about collective
attention, such as searches, reviews, and ratings, which now abound online, may
be used as indicators of what kinds of new ideas and innovations will ensue.

Whether there is a hard causal link between demand and supply remains an open
question. Our main contribution here has been to establish a quantitative
relation between the timing of demand and supply of information. A definition of
``information'' is more elusive than that of material goods; and quantifying
demand is particularly hard in this case. 

Our analysis focuses on aggregate-level behaviors. Models of individual browsing
behavior could shed more light on how people allocate their attention among
competing information stimuli online. Given the sensitive nature of the personal
information revealed by individual browsing habits, validating such models with 
data is a challenge, as revealed by the recent discussions about the
trade-offs between data-driven social science research and individual privacy
rights. Nevertheless, further empirical analyses and theoretical models of
individual and collective dynamics of attention will lead a better understanding
of the social exchange of knowledge in online and offline information networks.

\section*{Methods}

\subsection*{Data collection.}

In our analysis we used the public dataset generated by the servers of the
Wikimedia Foundation. Traffic volume is the number of non-unique HTTP requests
that an article receives, as a proxy for the popularity of the
subject \cite{Mestyan2013, Moat2013}. We collected data about hourly traffic to
the neighbors of Wikipedia articles created during 2012. The data were
pre-processed for analysis. We conflated titles that automatically redirect to
other entries. We used the information in the `redirect' table to perform this
check. We considered only pages created by humans, using a recent list of all
known bots to discard automatically-generated pages. Neighbors were found by
looking at the `pagelinks' table, after resolving redirects.
 
\subsection*{Page creation.}

To check whether a page was actually created during 2012 we consulted the time
stamp of its earliest recorded revision (the reference time stamp).
Unfortunately, this information is not always accurate since Wikipedia pages can
be merged, migrated, have their edit history fully or partially deleted, or even
lost. We thus checked that no traffic to the page had been recorded in our
dataset in a 50-week exclusion window before the reference time stamp. However,
because it is customary to include links to missing entries in order to
encourage other contributors to create them, we found this criterion to produce
too many false negatives. We settled for a small threshold, allowing pages with
at most 5\% (420) non-null hourly observations in the exclusion window.

\subsection*{Links.}

At its earliest stage a Wikipedia article rarely contains more than a handful of
sentences and links. As a consequence, looking at the early set of neighbors
would yield very sparse information. On the other hand, deletion of links is
rare \cite{Capocci2006}. Therefore we collected the neighbors that link to and
are linked by the page at the present day. 

\subsection*{Relative traffic change.}

Let us consider a focal page with $N$ neighbours and an observation window of
length $L$ centered around a reference time $t_c$, which is the time when the
page is created. The total traffic volume each neighbor receives before and
after $t_c$ corresponds to random variables $V^{(a)}_1, \ldots, V^{(a)}_N$ and
$V^{(b)}_1, \ldots, V^{(b)}_N$, respectively. The average volume change $\Delta
\overline V = \frac{1}{N}\sum_{i=1}^{N}V_i^{(a)} - V_i^{(b)}$ indicates whether,
on average, attention to a neighbor is more concentrated before the creation of
the page ($\Delta \overline V > 0$) or after ($\Delta \overline V < 0$).

Even though it accounts for the broad distribution of neighborhood sizes (see
Fig~2(a)), $\Delta \overline V$ does not guarantee a fair
comparison between topics for two reasons: first, the distribution of attention
across topics is broad (as shown in Fig.~3); second, Web
traffic is known to follow circadian, weekly, and seasonal
rhythms \cite{Thompson1997}. Over a week, an overall change in traffic volume
$\sum_i V_i^{(a)} - V_i^{(b)} = 10$ visits may represent a
dramatic surge of attention if observed over a group of pages that average $100$
visits per week. However, it would be barely noticeable if the same pages
averaged $10^{4}$ visits per week. To overcome these problems, let us define the
relative (median) traffic change: 
\begin{equation} \frac{\Delta V}{V} \equiv \frac{V^{(a)} - V^{(b)}}{V^{(a)} +
	V^{(b)}} \end{equation}
where $V^{(\cdot)}$ is the median traffic over a neighbor. We
choose to use the median since it is a more robust estimator in the
presence of outliers, and almost every article in our samples has at least one
very high-traffic neighbor (e.g., ``United States''), whose volume of traffic is
insensitive to all but the most high-profile events recorded in the dataset. We
also repeated our analysis using the sample mean and found qualitatively similar
results.

The length $L$ of the observation window must be chosen considering a trade-off
between competing requirements. Most attention spikes tend to be relatively
brief --- on the scale of the day --- and so the value of $L$ should not be too
large, to avoid lumping together consecutive attention bursts. On the
other hand, because of the strong circadian and weekly cycles that we see
in Fig.~1, $L$ cannot be too small, otherwise these
fluctuations would dominate the signal for all but the largest bursts. We
therefore consider a two-week observation window ($L = 14$ days), centered at
the time of creation of the new page. 

\subsection*{Baseline sample.}

To collect the baseline data we drew at random without replacement an existing
page (i.e., created before 2012) for each new page, and extracted traffic to its
neighbors at a random time stamp during 2012. We also repeated the analysis with
a different baseline sample, where instead of a random time stamp we used the
time of creation of the associated new page, and found similar results.

\subsection*{Acknowledgments} 

The authors would like to thank Richard Shiffrin, Filippo Radicchi, and
Yong-Yeol Ahn for useful suggestions, and John McCurley for his help with
revising the manuscript. We acknowledge the Wikimedia Foundation for making the
data available. This work was supported in part by the Swiss National Science
Foundation (fellowship 142353), NSF (grant CCF-1101743), the Lilly Endowment,
and the James S. McDonnell Foundation. The authors declare that they have no
competing financial interests. 

\subsection*{Author contributions} 

GLC, AF, and FM designed the research. GLC conducted data collection and
analysis. All authors prepared the manuscript. The authors declare no competing
financial interests.



\begin{thebibliography}{10}
\expandafter\ifx\csname url\endcsname\relax
  \def\url#1{\texttt{#1}}\fi
\expandafter\ifx\csname urlprefix\endcsname\relax\def\urlprefix{URL }\fi
\providecommand{\bibinfo}[2]{#2}
\providecommand{\eprint}[2][]{\url{#2}}

\bibitem{Crane2008}
\bibinfo{author}{Crane, R.} \& \bibinfo{author}{Sornette, D.}
\newblock \bibinfo{title}{Robust dynamic classes revealed by measuring the
  response function of a social system}.
\newblock \emph{\bibinfo{journal}{Proceedings of the National Academy of
  Sciences}} \textbf{\bibinfo{volume}{105}}, \bibinfo{pages}{15649--15653}
  (\bibinfo{year}{2008}).

\bibitem{Moat2013}
\bibinfo{author}{Moat, H.~S.} \emph{et~al.}
\newblock \bibinfo{title}{Quantifying {W}ikipedia usage patterns before stock
  market moves}.
\newblock \emph{\bibinfo{journal}{Sci. Rep.}} \textbf{\bibinfo{volume}{3}},
  \bibinfo{pages}{1801} (\bibinfo{year}{2013}).

\bibitem{Miritello2013}
\bibinfo{author}{Miritello, G.}, \bibinfo{author}{Lara, R.},
  \bibinfo{author}{Cebrian, M.} \& \bibinfo{author}{Moro, E.}
\newblock \bibinfo{title}{Limited communication capacity unveils strategies for
  human interaction}.
\newblock \emph{\bibinfo{journal}{Sci. Rep.}} \textbf{\bibinfo{volume}{3}}
  (\bibinfo{year}{2013}).

\bibitem{Weng2013}
\bibinfo{author}{Weng, L.} \emph{et~al.}
\newblock \bibinfo{title}{The role of information diffusion in the evolution of
  social networks}.
\newblock In \emph{\bibinfo{booktitle}{Proceedings of the 19th ACM SIGKDD
  International Conference on Knowledge Discovery and Data Mining}}, KDD '13,
  \bibinfo{pages}{356--364} (\bibinfo{publisher}{ACM}, \bibinfo{address}{New
  York, NY, USA}, \bibinfo{year}{2013}).

\bibitem{Benkler2006}
\bibinfo{author}{Benkler, Y.}
\newblock \emph{\bibinfo{title}{The wealth of networks: How social production
  transforms markets and freedom}} (\bibinfo{publisher}{Yale University Press},
  \bibinfo{year}{2006}).

\bibitem{Shapiro1999}
\bibinfo{author}{Shapiro, C.} \& \bibinfo{author}{Varian, H.~R.}
\newblock \emph{\bibinfo{title}{Information Rules: A Strategic Guide to the
  Network Economy}} (\bibinfo{publisher}{Harvard Business Press},
  \bibinfo{year}{1999}).

\bibitem{Simon1971}
\bibinfo{author}{Simon, H.~A.}
\newblock \emph{\bibinfo{title}{Computers, communications, and the public
  interest}}, vol.~\bibinfo{volume}{72}, chap. \bibinfo{chapter}{Designing
  organizations for an information-rich world}, \bibinfo{pages}{37}
  (\bibinfo{publisher}{Johns Hopkins Press, Baltimore, MD},
  \bibinfo{year}{1971}).

\bibitem{Wu2007}
\bibinfo{author}{Wu, F.} \& \bibinfo{author}{Huberman, B.~A.}
\newblock \bibinfo{title}{Novelty and collective attention}.
\newblock \emph{\bibinfo{journal}{Proceedings of the National Academy of
  Sciences}} \textbf{\bibinfo{volume}{104}}, \bibinfo{pages}{17599--17601}
  (\bibinfo{year}{2007}).

\bibitem{Hodas2012}
\bibinfo{author}{Hodas, N.~O.} \& \bibinfo{author}{Lerman, K.}
\newblock \bibinfo{title}{How visibility and divided attention constrain social
  contagion}.
\newblock In \emph{\bibinfo{booktitle}{Privacy, Security, Risk and Trust
  (PASSAT), 2012 International Conference on and 2012 International Conference
  on Social Computing (SocialCom)}}, \bibinfo{pages}{249--257}
  (\bibinfo{organization}{IEEE}, \bibinfo{year}{2012}).

\bibitem{Weng2012}
\bibinfo{author}{Weng, L.}, \bibinfo{author}{Flammini, A.},
  \bibinfo{author}{Vespignani, A.} \& \bibinfo{author}{Menczer, F.}
\newblock \bibinfo{title}{Competition among memes in a world with limited
  attention}.
\newblock \emph{\bibinfo{journal}{Sci. Rep.}} \textbf{\bibinfo{volume}{2}},
  \bibinfo{pages}{335} (\bibinfo{year}{2012}).

\bibitem{Ginsberg2009}
\bibinfo{author}{Ginsberg, J.} \emph{et~al.}
\newblock \bibinfo{title}{Detecting influenza epidemics using search engine
  query data}.
\newblock \emph{\bibinfo{journal}{Nature}} \textbf{\bibinfo{volume}{457}},
  \bibinfo{pages}{1012--1014} (\bibinfo{year}{2009}).

\bibitem{Mestyan2013}
\bibinfo{author}{Mesty{\'a}n, M.}, \bibinfo{author}{Yasseri, T.} \&
  \bibinfo{author}{Kert{\'e}sz, J.}
\newblock \bibinfo{title}{Early prediction of movie box office success based on
  {W}ikipedia activity big data}.
\newblock \emph{\bibinfo{journal}{PLoS ONE}} \textbf{\bibinfo{volume}{8}},
  \bibinfo{pages}{e71226} (\bibinfo{year}{2013}).

\bibitem{Leskovec2009}
\bibinfo{author}{Leskovec, J.}, \bibinfo{author}{Backstrom, L.} \&
  \bibinfo{author}{Kleinberg, J.}
\newblock \bibinfo{title}{Meme-tracking and the dynamics of the news cycle}.
\newblock In \emph{\bibinfo{booktitle}{Proceedings of the 15th ACM SIGKDD
  International Conference on Knowledge Discovery and Data Mining}}, KDD '09,
  \bibinfo{pages}{497--506} (\bibinfo{publisher}{ACM}, \bibinfo{address}{New
  York, NY, USA}, \bibinfo{year}{2009}).

\bibitem{Simmons2011}
\bibinfo{author}{Simmons, M.~P.}, \bibinfo{author}{Adamic, L.~A.} \&
  \bibinfo{author}{Adar, E.}
\newblock \bibinfo{title}{Memes online: Extracted, subtracted, injected, and
  recollected.}
\newblock In \emph{\bibinfo{booktitle}{Proceedings of the Fifth International
  AAAI Conference on Weblogs and Social Media}} (\bibinfo{year}{2011}).

\bibitem{Huberman2009}
\bibinfo{author}{Huberman, B.~A.}, \bibinfo{author}{Romero, D.~M.} \&
  \bibinfo{author}{Wu, F.}
\newblock \bibinfo{title}{Crowdsourcing, attention and productivity}.
\newblock \emph{\bibinfo{journal}{Journal of Information Science}}
  \textbf{\bibinfo{volume}{35}}, \bibinfo{pages}{758--765}
  (\bibinfo{year}{2009}).

\bibitem{Franck1999}
\bibinfo{author}{Franck, G.}
\newblock \bibinfo{title}{Scientific communication--a vanity fair?}
\newblock \emph{\bibinfo{journal}{Science}} \textbf{\bibinfo{volume}{286}},
  \bibinfo{pages}{53--55} (\bibinfo{year}{1999}).

\bibitem{Kampf2012}
\bibinfo{author}{K{\"a}mpf, M.}, \bibinfo{author}{Tismer, S.},
  \bibinfo{author}{Kantelhardt, J.~W.} \& \bibinfo{author}{Muchnik, L.}
\newblock \bibinfo{title}{Fluctuations in {W}ikipedia access-rate and
  edit-event data}.
\newblock \emph{\bibinfo{journal}{Physica A: Statistical Mechanics and its
  Applications}} \textbf{\bibinfo{volume}{391}}, \bibinfo{pages}{6101--6111}
  (\bibinfo{year}{2012}).

\bibitem{Ratkiewicz2010a}
\bibinfo{author}{Ratkiewicz, J.}, \bibinfo{author}{Fortunato, S.},
  \bibinfo{author}{Flammini, A.}, \bibinfo{author}{Menczer, F.} \&
  \bibinfo{author}{Vespignani, A.}
\newblock \bibinfo{title}{Characterizing and modeling the dynamics of online
  popularity}.
\newblock \emph{\bibinfo{journal}{Phys. Rev. Lett.}}
  \textbf{\bibinfo{volume}{105}}, \bibinfo{pages}{158701}
  (\bibinfo{year}{2010}).

\bibitem{Lehmann2012}
\bibinfo{author}{Lehmann, J.}, \bibinfo{author}{Gon\c{c}alves, B.},
  \bibinfo{author}{Ramasco, J.~J.} \& \bibinfo{author}{Cattuto, C.}
\newblock \bibinfo{title}{Dynamical classes of collective attention in
  {T}witter}.
\newblock In \emph{\bibinfo{booktitle}{Proceedings of the 21st International
  Conference on World Wide Web}}, WWW '12, \bibinfo{pages}{251--260}
  (\bibinfo{publisher}{ACM}, \bibinfo{address}{New York, NY, USA},
  \bibinfo{year}{2012}).

\bibitem{Goncalves2008}
\bibinfo{author}{Gon\c{c}alves, B.} \& \bibinfo{author}{Ramasco, J.~J.}
\newblock \bibinfo{title}{Human dynamics revealed through web analytics}.
\newblock \emph{\bibinfo{journal}{Phys. Rev. E}} \textbf{\bibinfo{volume}{78}},
  \bibinfo{pages}{026123} (\bibinfo{year}{2008}).

\bibitem{Meiss2008}
\bibinfo{author}{Meiss, M.~R.}, \bibinfo{author}{Menczer, F.},
  \bibinfo{author}{Fortunato, S.}, \bibinfo{author}{Flammini, A.} \&
  \bibinfo{author}{Vespignani, A.}
\newblock \bibinfo{title}{Ranking web sites with real user traffic}.
\newblock In \emph{\bibinfo{booktitle}{Proceedings of the 2008 International
  Conference on Web Search and Data Mining}}, WSDM '08, \bibinfo{pages}{65--76}
  (\bibinfo{publisher}{ACM}, \bibinfo{address}{New York, NY, USA},
  \bibinfo{year}{2008}).

\bibitem{Alexa.com}
\bibinfo{author}{Alexa.com}.
\newblock \bibinfo{title}{Alexa top 500 global sites}.
\newblock \bibinfo{howpublished}{\url{http://www.alexa.com/topsites}}
  (\bibinfo{year}{2014}).
\newblock \bibinfo{note}{Last visited January 2014}.

\bibitem{Google2012}
\bibinfo{author}{Google}.
\newblock \bibinfo{title}{Zeitgest 2012 -- {G}oogle}.
\newblock \bibinfo{howpublished}{\url{http://www.google.com/zeitgeist/2012/}}
  (\bibinfo{year}{2012}).
\newblock \bibinfo{note}{Last visited January 2014}.

\bibitem{Ratkiewicz2010}
\bibinfo{author}{Ratkiewicz, J.}, \bibinfo{author}{Flammini, A.} \&
  \bibinfo{author}{Menczer, F.}
\newblock \bibinfo{title}{Traffic in social media {I}: Paths through
  information networks}.
\newblock In \emph{\bibinfo{booktitle}{Social Computing (SocialCom), 2010 IEEE
  Second International Conference on}}, \bibinfo{pages}{452--458}
  (\bibinfo{year}{2010}).

\bibitem{Mas-Colell1995}
\bibinfo{author}{Mas-Colell, A.}, \bibinfo{author}{Whinston, M.~D.},
  \bibinfo{author}{Green, J.~R.} \emph{et~al.}
\newblock \emph{\bibinfo{title}{Microeconomic theory}},
  vol.~\bibinfo{volume}{1} (\bibinfo{publisher}{Oxford University Press New
  York}, \bibinfo{year}{1995}).

\bibitem{WikimediaFoundation2014}
\bibinfo{author}{{Wikimedia Foundation}}.
\newblock \bibinfo{title}{Wikipedia statistics - tables - active wikipedians}.
\newblock
  \bibinfo{howpublished}{\url{http://stats.wikimedia.org/EN/TablesWikipediansEditsGt5.htm}}
  (\bibinfo{year}{2014}).
\newblock \bibinfo{note}{Last visited July 2014}.

\bibitem{Capocci2006}
\bibinfo{author}{Capocci, A.} \emph{et~al.}
\newblock \bibinfo{title}{Preferential attachment in the growth of social
  networks: The {I}nternet encyclopedia {W}ikipedia}.
\newblock \emph{\bibinfo{journal}{Phys. Rev. E}} \textbf{\bibinfo{volume}{74}},
  \bibinfo{pages}{036116} (\bibinfo{year}{2006}).

\bibitem{Thompson1997}
\bibinfo{author}{Thompson, K.}, \bibinfo{author}{Miller, G.~J.} \&
  \bibinfo{author}{Wilder, R.}
\newblock \bibinfo{title}{Wide-area {I}nternet traffic patterns and
  characteristics}.
\newblock \emph{\bibinfo{journal}{{IEEE} Network: Mag. of Global Internetwkg.}}
  \textbf{\bibinfo{volume}{11}}, \bibinfo{pages}{10--23}
  (\bibinfo{year}{1997}).

\bibitem{Geipel2007}
\bibinfo{author}{Geipel, M.~M.}
\newblock \bibinfo{title}{Self-organization applied to dynamic network layout}.
\newblock \emph{\bibinfo{journal}{International Journal of Modern Physics C}}
  \textbf{\bibinfo{volume}{18}}, \bibinfo{pages}{1537--1549}
  (\bibinfo{year}{2007}).

\end{thebibliography}

\newpage

\begin{table}[ht]
  \centering
  \caption{Top Wikipedia entries by relative traffic change in absolute value
    $\left|\Delta V/V\right|$. We consider articles whose neighbors received at
    least $5\times 10^5$ visits per day, and report the total traffic
    $V_\textrm{tot}$ within the entire 
  two-week window.} 
\begin{tabular}{lrr}
\toprule
                                                    Article & $V_\textrm{tot}$ ($\times 10^6$ visits) & $\Delta V/V$ \\
\midrule
                                         Katherine Copeland &                     $8.66$ &        $+0.76$ \\
                        A Symphony of British Music (album) &                    $10.06$ &        $+0.69$ \\
                                  Elizabeth Price (gymnast) &                     $7.23$ &        $+0.65$ \\
                                Operation Pillar of Defense &                    $18.63$ &        $+0.49$ \\
                                               Lin Qingfeng &                     $8.08$ &        $+0.44$ \\
                                     J. Christopher Stevens &                     $7.84$ &        $+0.44$ \\
                                       2013 Australian Open &                    $12.06$ &        $+0.43$ \\
                                         Niluka Karunaratne &                     $8.75$ &        $+0.42$ \\
                                                   Li Yunqi &                     $8.35$ &        $+0.42$ \\
                                                  Kony 2012 &                    $20.42$ &        $+0.36$ \\
\dots & \dots & \dots \\
        United States presidential election in Idaho, 2012 &                     $7.88$ &        $-0.22$ \\
      United States presidential election in Vermont, 2012 &                     $7.80$ &        $-0.23$ \\
     United States presidential election in Oklahoma, 2012 &                     $7.88$ &        $-0.24$ \\
 United States presidential election in Rhode Island, 2012 &                     $7.70$ &        $-0.25$ \\
     United States presidential election in Maryland, 2012 &                     $7.02$ &        $-0.25$ \\
     United States presidential election in Illinois, 2012 &                     $8.47$ &        $-0.26$ \\
    United States presidential election in Tennessee, 2012 &                     $8.06$ &        $-0.26$ \\
                                     2012 BNP Paribas Open &                    $11.90$ &        $-0.34$ \\
                                              Rhoda Abbott &                     $7.17$ &        $-0.36$ \\
                                             Kelley Hurley &                     $8.09$ &        $-0.51$ \\
\bottomrule
\end{tabular}
\label{tab:topbursts}
\end{table}

\begin{figure}[ht]
  \centering
  \includegraphics{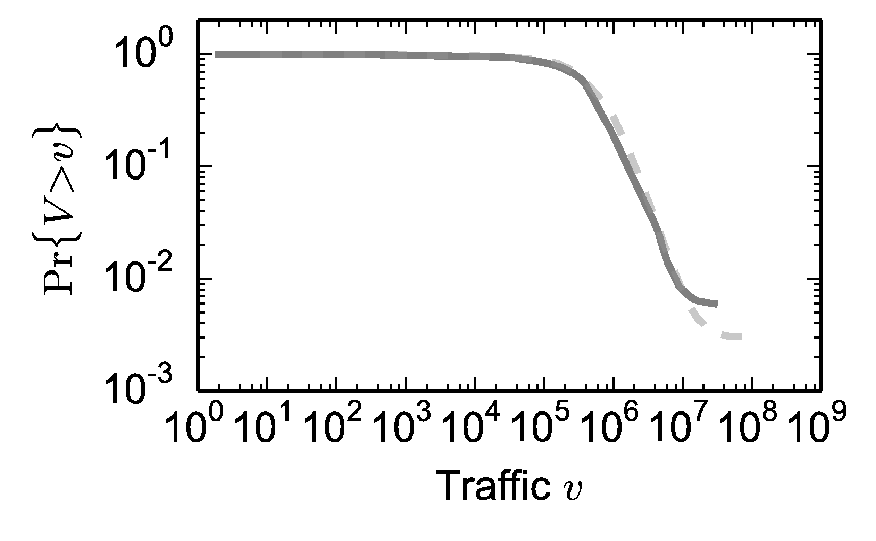}
  \linespread{1.5}
  \caption{\textbf{Distribution of traffic volume}. Cumulative distribution of
  total traffic volume $V$ for articles created in 2012 (solid) and pre-existing
2012 (dashed).} 
\label{fig:attentiondist}
\end{figure}

\end{document}